\documentclass[twocolumn]{revtex4}
\usepackage{graphicx}% Include figure files

\newcommand{\beq}{\begin{equation}}
\newcommand{\eeq}{\end{equation}}
\newcommand {\brho} {\mbox{\boldmath$\rho$}}

\begin{document}

\title{Comparison of the DFT and field approaches to Van der Waals interactions
in planparallel geometry}
\author{Gregor Veble}
%\email{gregor.veble@fmf.uni-lj.si}
\altaffiliation[Also at ]{Center for Applied Mathematics and Theoretical Physics,
University of Maribor, Maribor, Slovenia}
\author{Rudolf Podgornik}
%\email{rudolf.podgornik@fmf.uni-lj.si}
\altaffiliation[Also at ]{Department  of Theoretical Physics,
J. Stefan Institute, Ljubljana, Slovenia}
\affiliation{Department of Physics, Faculty of Mathematics and Physics, \\
University of Ljubljana,  Jadranska 19, SI-1000 Ljubljana,  Slovenia}
\date{\today}

\begin{abstract}
We establish a general equivalence between van der Waals interaction energies within the formalism
of the non-local van der Waals functional of the density functional theory and within the formalism
of the field approach based on the  secular determinants of the electromagnetic field modes. We then compare the two methods explicitly in the case of a planparallel geometry with a continuously varying dielectric response function and show that their respective {\sl numerical implementations} are not equivalent. This allows us to discuss the merits of the two approaches and possible advantages of either method in a simple model calculation.
\end{abstract}

\maketitle

\section{Introduction}

Van der Waals (vdW) interactions are ubiquitous and their effects are quintessential for understanding various properties of physical, chemical and biological systems \cite{parsegian}.
In particular the planparallel layer geometry has played a prominent role in our understanding of  phospholipid \cite{rudi1} and  polymer assemblies  \cite{decher1,sukhishvili}, as well as in inorganic systems such as the intergranular films in silicon nitride structural ceramics \cite{roger-1} or interfaces and grain boundaries in perovskite based  electronic ceramics \cite{roger-2}. Understanding molecular
interactions in these systems is an important step in controlling the assembly
process. Though  interactions in these assemblies are due to many different
specific  properties, vdW  interactions are a common underlying and unifying feature.

The vdW interaction energy (or free energy if temperature effects are of relevance) can be derived in many different ways \cite{parsegian,ninham} which are in principle equivalent but vary in terms of their implementability for specific problems and geometries. Though all these approaches differ in various ways, they are all based on considerations of the electromagnetic field equations and the changes in the configurations of these fields when bodies are brought into close proximity.

A different approach to vdW interactions stemms from the application of density functional theory (DFT)  
\cite{parr} that has been very successful in the study of individual molecules as well as dense solids. In this framework the vdW interactions belong to the non-local effects that need to be considered when applying DFT to sparse matter \cite{rydberg-dion}. It is now well recognized 
\cite{langreth} that long-range effects in the DFT are not accounted for by the local or semilocal approximations of the exchange-correlation functional, though the vdW interaction is indeed a long-range correlation effect that is nevertheless unrelated to exchange correlations. A large body of work 
(see \cite{rydberg-dion,langreth,rydberg-dion2,dobson} and references therein) has thus been devoted to a seamless extension of the DFT energy functionals that would properly capture the saturation of the vdW interaction at small separation and its continuous approach to the standard DFT short-range interaction energy. In this context the vdW interactions in the planparallel slab geometry are derived {\sl via} the matter response functions and successfully capture the seamless transition of long-range vdW interactions to those important at near contact conditions \cite{dobson}. 

In this work we will show that formulations of vdW interactions based on the field point of view \cite{reform} and on the matter point of view \cite{rydberg-dion} give the same interaction energy in the planparallel slab geometry with continuously varying dielectric response. This is just a corrolary on the general equivalence between the field and the matter approaches to the vdW interactions. Apart from that we will also show that the two approaches are however not equivalent when it comes to numerical implementation of the calculation of the vdW interactions, where the field approach seems to be numerically advantageous.

\section{Description of the problem}

We first present the expression for the vdW interaction energy derived within the DFT approach and the field approach and later show their equivalence. In what follows we will limit ourselves to the nonretarded limit of a continuously (with a continuous derivative) varying local dielectric profile in the coordinate perpendicular to the two apposed planar surfaces at zero temperature. Therefore we can write for the dielectric response function
\beq
\epsilon ({\bf r}, {\bf r }'; i\xi) = \delta(z -z') \int \frac{d^2 \bf Q}{(2\pi)^2} ~\epsilon(Q; z; i\xi) ~e^{ i {\bf Q} \cdot ({\brho} - {\brho}')},
\label{eq:eps}
\eeq
where $\bf Q$ is the wavevector perpendicular to the axis $z$, $\brho$ is the 2D radius vector in the plane perpendicular to $z$ axis, $i \xi$ is the imaginary frequency with the dielectric function of imaginary frequencies being real and decreasing with increasing frequency \cite{parsegian}. In what follows we shall use $\epsilon(z)$ as a shorthand notation for $\epsilon(Q; z; i\xi) $.

\begin{figure}
\centerline{\includegraphics[width=8cm]{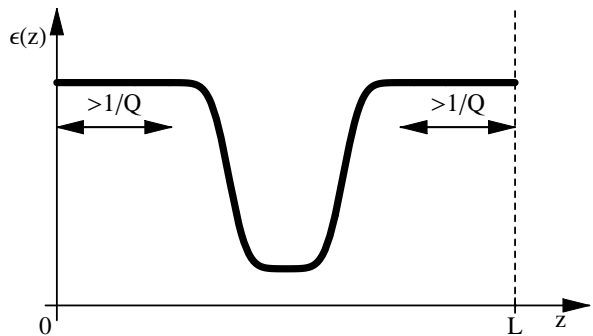}}
\caption{A schematic diagram of a typical dielectric profile (arbitrary scale) and the interval $[0,L]$ as required by the DFT approach. Near both ends of the interval the dielectric response should be constant for about at least $\propto 1/Q$. \label{fig:diagram}}
\end{figure}

Within the DFT formalism proposed by Rydberg {\sl et al.} \cite{rydberg-dion,rydberg-dion2} the vdW interaction energy can be obtained  as
\beq
F/A=\int_0^\infty \frac{d \xi}{2 \pi} \int \frac{d^2Q}{(2 \pi)^2} \ln \frac{D_M(i\xi; L)}{D_M^{(0)}(i\xi)},
\label{eq:freeen}
\eeq
where the expression $D_M(i\xi; L)$ referrs to the two planar interfaces at separation $L$, while $D_M^{(0)}(i\xi)$ referrs to a reference system (usually empty space).  In the zero-temperature limit, the computation of $D_M(i\xi; L)$ and $D_M^{(0)}(i\xi)$ can be reduced  \cite{rydberg-dion,rydberg-dion2} to the solution of the modified scalar Laplace equation
\beq
\tilde \varphi^{\prime\prime}+\frac{\epsilon^\prime}{\epsilon} \tilde \varphi^\prime - Q^2 \tilde \varphi=0.
\label{eq:determinantphi}
\eeq
if one identifies 
\beq
D_M(i\xi; L) = (\tilde \varphi^\prime(0))^{-1}, ~\qquad{\rm with~boundary~conditions} \qquad
\tilde \varphi(0)=0,\ \tilde \varphi(L)=1.
\label{eq:ddet}
\eeq
The origin and the distance $L$ need to be chosen such that the dielectric constant does not vary sufficiently close to both edges of the interval (within a distance $\propto 1/Q$), as is shown in Figure
\ref{fig:diagram}.

In the field approach the calculation of the vdW interaction energy can be reduced to an algebra of $2\times 2$ matrices \cite{reform}. The interaction energy still retains the form
\beq
F/A=\int_0^\infty \frac{d \xi}{2 \pi} \int \frac{d^2Q}{(2 \pi)^2} \ln \frac{D_F(i\xi; L)}{D_F^{(0)}(i\xi)},
\label{eq:freeen-1}
\eeq
but here the interpretation of  $D(i\xi; L)$ is different. In fact one derives that $D_F(i\xi; L) = M_{11}$ is the $(1,1)$ matrix element of the product of  matrices 
\beq
M=\ldots  T_{i+1} D_{i+1} T_i D_i \ldots, \label{eq:matrixprod}
\eeq
where the index $i$ runs over the whole interval of properly discretized $z$ axis, defined as
\beq
T_i=\left[
\begin{array}{cc}
1 & 0 \\
0 & \exp(-2 \rho_i \delta z_i)\\
\end{array}
\right], \ \ 
D_i=\left[
\begin{array}{cc}
1 & -\Delta_i\\
-\Delta_i & 1\\
\end{array}
\right],
\eeq
with $\delta z_i$ the slab thickness, {\sl i.e.} the separation between the $i-th$ and $(i+1)-th$ layers.
Here $\rho_i=\sqrt{Q^2+\epsilon_i \xi^2/c^2}$, and the quantities describing the relative  dielectric missmatch at each layer $i$ are defined as
 $$\Delta_i=\frac{\rho_{i+1} \epsilon_i -\rho_i \epsilon_{i+1}}
{\rho_{i+1} \epsilon_i+\rho_i \epsilon_{i+1}}$$for TM  and
$$\Delta_i=\frac{\rho_{i+1}  -\rho_i}{\rho_{i+1}+\rho_i}$$ for TE modes, where we assume that the magnetic permeability equals $1$ in the whole spatial domain. $T_i$ matrix corresponds to the phase shift of the left and right travelling waves when 
moving across the $i$-th slab, and $D_i$ matrix links the fields at the boundaries 
between slabs. The matrices $T$ and $D$ are strictly real.

Off hand it is not clear whether the two expressions, {\sl i.e.} (\ref{eq:freeen}) and (\ref{eq:freeen-1}), for the vdW interaction energy are the same, since the definitions of $D(i\xi; L)$ in both cases are not obviously related. In what follows we will first of all show that the two expressions are not only related but are, modulo some unimportant scaling factors, in fact exactly the same.

\section{Proof of equivalence}

We now investigate whether the two approaches to the vdW interaction energy give formally the same 
result. We start with equation (\ref{eq:determinantphi}) from the DFT approach and note that the solution of any equation of the form
\beq
y^{\prime\prime}+\beta(z) y^\prime - Q^2(z) y=0
\eeq
can be written as
\beq
y=y_1+y_2,
\eeq
where $y_1$ and $y_2$ satisfy a system of two coupled linear differential equations
\beq
\left[\begin{array}{c}
y_1^\prime\\
y_2^\prime
\end{array}\right]=
\left[\begin{array}{cc}
Q(z) - \gamma(z) & \gamma(z)\\
\gamma(z) & -Q(z)-\gamma(z)
\end{array}\right]
\left[\begin{array}{c}
y_1\\
y_2
\end{array}\right], 
\label{eq:splitdet}
\eeq
with
\beq
\gamma(z)=\frac{\beta(z)+\frac{Q^\prime(z)}{Q(z)}}{2}.
\eeq
In the DFT approach, as $Q$ does not depend on $x$, we have 
\beq
\gamma(z)=\frac{1}{2} \left(\ln (\epsilon)\right)^\prime.
\eeq

We will now try and reformulate the matrix product (\ref{eq:matrixprod}) in the field approach so that it will lead to the same equation (\ref{eq:splitdet}) as the DFT approach. The matrix approach can be somewhat simplified by noting that the initial condition for the matrix  $M$ can be written as 
\beq
M_0=\left[
\begin{array}{cc}
1 & 0\\
0 & 0\\
\end{array}
\right]=
\left[
\begin{array}{c}
1\\
0
\end{array}
\right]
\otimes
\left[
\begin{array}{cc}
1 & 0
\end{array}
\right].
\eeq
This is due to the fact that, at nonzero $\rho$, the product of matrices $T_i$ for the homogeneous space extending to $-\infty$ will converge to the above matrix, and thus it can be taken as the initial condition for the matrix $M$ at any coordinate $z_0$, where for $z<z_0$ the dielectric constant is nonvarying. Such splitting of the matrix reduces the problem of the matrix product to a problem of  matrices acting on a vector, in the sense that after $i$ layers
\beq
\left[\begin{array}{c}
\psi_1^{(i)}\\
\psi_2^{(i)}
\end{array}\right]=T_{i} D_{i} T_{i-1} D_{i-1} \ldots T_1 D_1
\left[\begin{array}{c}
1\\
0
\end{array}\right].
 \label{eq:psiproduct}
\eeq
At the final layer we may identify $\psi_1=M_{11}$, $\psi_2=M_{21}$ and therefore
\beq
D_{F}=\psi_1.
\label{eq:dmat}
\eeq 
Let us now reduce the matrix product in the field formulation into a set of coupled differential equations for a continuously varying dielectric profile. Here we will treat only the nonretarded case, $c \longrightarrow \infty$,  where $\rho=Q$.
If the layer thickess is small, both the matrix $T_i$ as well as $D_i$ differ only slightly from the identity, such that
\beq
T_i\approx I+\left[
\begin{array}{cc}
0 & 0 \\
0 & -2 Q \\
\end{array}
\right]\delta z_i,
\\ \ D_i\approx I+
\left[
\begin{array}{cc}
0 & -\gamma_i \\
-\gamma_i & 0\\
\end{array}
\right]\delta z_i,
\eeq
where $\gamma_i=\Delta_i/\delta z_i$. This is a difference scheme that in the limit $\delta z_i\to 0$ leads to a continuous formulation of the equation (\ref{eq:psiproduct}),
\beq
\left[\begin{array}{c}
\psi_1^\prime\\
\psi_2^\prime
\end{array}\right]=
\left[\begin{array}{cc}
0 & -\gamma(z)\\
-\gamma(z) & -2 Q
\end{array}\right]
\left[\begin{array}{c}
\psi_1\\
\psi_2
\end{array}\right],
\label{eq:splitdet-1}
\eeq
with the initial condition
$\psi_1(-\infty)=1$, $\psi_2(-\infty)=0$. 
Let us now transform the system of differential equations (\ref{eq:splitdet-1}) obtained via the  matrix product calculation so that it will coincide with the system (\ref{eq:splitdet}) as derived from the DFT approach.
Obviously eqs. (\ref{eq:splitdet}) and (\ref{eq:splitdet-1}) differ in the diagonal elements of the matrix. However, if we introduce the transformation
\begin{eqnarray}
\bar\psi_i(z)=\exp\left(\int_{0}^z (Q-\gamma(z^\prime))dz^\prime\right)  \psi_i(z) =\nonumber \\
= \sqrt{\frac{\epsilon(0)}{\epsilon(z)}} \ \exp(Qz)\  \psi_i(z) 
\label{eq:diagtrans}
\end{eqnarray}
we then derive the following system of equations
\beq
\left[\begin{array}{c}
\bar\psi_1^\prime\\
\bar\psi_2^\prime
\end{array}\right]=
\left[\begin{array}{cc}
Q-\gamma(z) & -\gamma(z)\\
-\gamma(z) & -Q-\gamma(z)
\end{array}\right]
\left[\begin{array}{c}
\bar\psi_1\\
\bar\psi_2
\end{array}\right].
\eeq
This is still not equal to eq. (\ref{eq:splitdet}). The next transformation necessary is
\beq
\tilde \psi_1=\bar \psi_1,\ \tilde \psi_2=-\bar \psi_2, 
\label{eq:signtrans}
\eeq 
which now yields finally 
\beq
\left[\begin{array}{c}
\tilde \psi_1^\prime\\
\tilde \psi_2^\prime
\end{array}\right]=
\left[\begin{array}{cc}
Q-\gamma(z) & \gamma(z)\\
\gamma(z) & -Q-\gamma(z)
\end{array}\right]
\left[\begin{array}{c}
\tilde \psi_1\\
\tilde \psi_2
\end{array}\right].
\eeq
Obviously now the two sets of equations (\ref{eq:splitdet}) and (\ref{eq:splitdet-1}) coincide, which is the main step towards demonstrating the equivalence. Note also that the transformation
\begin{eqnarray}
\chi_i(z)=\exp\left(\int_{0}^z (Q+\gamma(z^\prime))dz^\prime\right)  \psi_i(z)= \nonumber\\
=\sqrt{\frac{\epsilon(z)}{\epsilon(0)}} \ \exp(Qz)\  \psi_i(z)
\end{eqnarray}
yields the system of equations 
\beq
\left[\begin{array}{c}
\chi_1^\prime\\
\chi_2^\prime
\end{array}\right]=
\left[\begin{array}{cc}
Q+\gamma(z) & -\gamma(z)\\
-\gamma(z) & -Q+\gamma(z)
\end{array}\right]
\left[\begin{array}{c}
\chi_1\\
\chi_2
\end{array}\right]
\eeq
According to equation (\ref{eq:splitdet}), the sum $H_\perp=\chi_1+\chi_2$ then satisfies the 
differential equation
\beq
\tilde H_\perp^{\prime\prime}-\frac{\epsilon^\prime}{\epsilon} \tilde H_\perp^\prime - Q^2 H_\perp=0
\label{eq:hperp}
\eeq
and describes the TM magnetic field modes. The only difference w.r.t. equation (\ref{eq:determinantphi}) is the sign in front of the term containing the first derivative, but we will show that this sign difference does not affect the interaction energy computation.

We thus  need to show that the expressions (\ref{eq:ddet}) and (\ref{eq:dmat}) for $D_{F}$ and $D_{M}$ as used in the energy integral in equations (\ref{eq:freeen}) and (\ref{eq:freeen-1}) give the same result. 
Let us rewrite the energy expression in the field approach as
\beq
\ln\frac{D_{F}(L)}{D_{F}^{(0)}(L)}=
\ln\frac{\psi_1(L)}{\psi_1^{(0)} (L) }=
\ln\frac{\psi_1(L) ~\psi_1^{(0)}(0)}{\psi_1(0)~ \psi_1^{(0)} (L)}. \label{eq:involved}
\eeq
Since the additional terms $\psi_1(0)$, $\psi_1^{(0)}(0)$ are equal to $1$ with the usual choice of initial conditions, due to the linearity of the problem, the above expression allows for a scaling of the solution without changing the result. 

It now helps to think about the DFT approach in terms of the split system of equations (\ref{eq:splitdet}), where $\tilde \varphi=y_1+y_2$. As the second component $y_2$ is exponentially decreasing, when $L$ is sufficiently large, we may write
$\tilde \varphi(L)\approx y_1$ and therefore $y_1(L)=1$. For the boundary condition $\tilde \varphi(0)=0$ to be fulfilled, we must require $y_1(0)=-y_2(0)$. This is, of course, a different boundary condition than in the matrix approach, where $\psi_1(0)=1$, $\psi_2(0)=0$. Since, however,  in a homogeneous region the second component decays exponentially fast, and if the homogeneous region extends sufficiently far into the $z>0$ region, the two solutions are largely equivalent when rescaled properly. According to equation (\ref{eq:splitdet}), the derivative that enters the free energy computation is then $\tilde \varphi^\prime(0)=2 Q ~ y_1(0)$. It then follows that
\begin{eqnarray}
\ln\frac{D_{M}(L)}{D_{M}^{(0)}(L)}=
\ln\frac{\tilde \varphi^{(0)\prime}(0)}{\tilde \varphi^\prime(0) }
=\ln \frac{2Q~ y_1^{(0)}(0)}{2Q~ y_1(0)}=\nonumber \\
=
\ln \frac{y_1(L)~y_1^{(0)}(0)}{y_1(0)~y_1^{(0)}(L)}, \label{eq:involveddet}
\end{eqnarray}
which agrees with the expression (\ref{eq:involved}) save for the transformations (\ref{eq:diagtrans}) and (\ref{eq:signtrans}) which take us from the variables $\psi_i$ to $y_i$. 

To show that these transformations do not affect the result, let us now assume that the dielectric functions $\epsilon(z)$ for the system at study and the reference system have the same limit at $z\to \infty$, while at the same time they also both limit to a possibly different value at $z\to -\infty$. If the transformations  (\ref{eq:diagtrans}, \ref{eq:signtrans}) are then applied to the expression (\ref{eq:involved}), the effects of the transformations exactly cancel for the original and reference systems and we are left with the expression (\ref{eq:involveddet}), concluding the complete  demonstration of equivalence.

\section{Consequences and discussion}

Though we have just shown that in principle the DFT and the field  approaches are equivalent, there are subtle differences in the numerical implementation that allow for a sensible comparison of the two methods which we discuss below.

First we would like to comment on the Lifshitz limit, where the dielectric profile consists of two semi-infinite halfspaces with one dielectric constant, separated by a finite slab of thickness $d$ with a different dielectric constant. While the matrix product approach  can deal with problems of this type  directly, the continuous picture of either the field  or the DFT approach encounters difficulties. The first issue is the definition of the derivative
\beq
\gamma(z)=\frac{1}{2} (\ln (\epsilon))^\prime=\frac{\epsilon^\prime}{2 \epsilon}
\eeq 
at the dielectric boundary. While for an interface at $z=0$ and having a jump in the dielectric constant 
$\Delta \epsilon$ we may write $\epsilon^\prime=\Delta \epsilon~ \delta(z)$, it is the value of $\epsilon$ in the denominator of the above equation that is not well defined at the interface. Furthermore, equations
(\ref{eq:hperp}) and (\ref{eq:determinantphi}) put $\gamma$  and therefore the $\delta(z)$ function in front of the first order derivative term, making it impossible to establish the proper connection formulae without further physical insight. This insight is, however, already incorporated in the discrete version of the matrix product formulation in which the Lifshitz case is handled gracefully and emerges naturally.

\begin{figure}[h]
\centerline{\includegraphics[width=8cm]{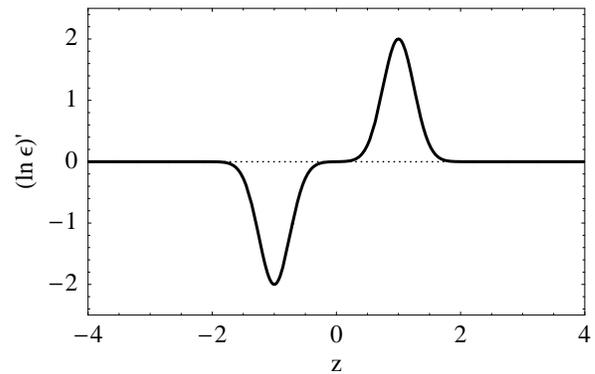}}
\caption{Plot of $\left(\ln \epsilon\right)^\prime$ for the model  dielectric response function, equation (\ref{enacba-28}), for the values of parameters parameters $a=1$, $\sigma=1/4$ and $A=2$.\label{fig:diel}}
\end{figure}
\begin{figure}[h]
\centerline{\includegraphics[width=8cm]{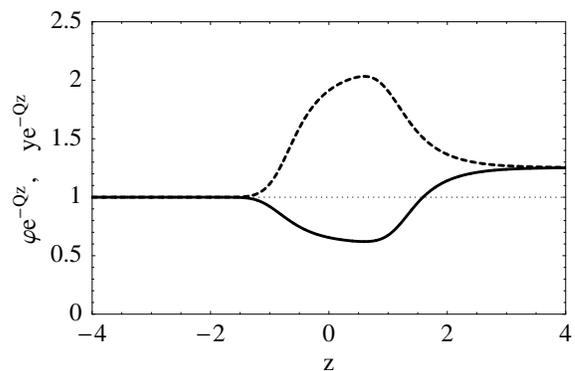}}
\caption{Comparison of the solution of the two differential equations (\ref{eq:hperp}) and (\ref{eq:determinantphi}) that only differ in the sign of the first order derivative term. In both cases, the result was divided by $\propto \exp(Qz)$. The full line shows the positive sign as in equation (\ref{eq:determinantphi}), the dashed line the negative one as in equation (\ref{eq:hperp}). The dotted line shows the value of $1$.\label{fig:seconddiff}}
\end{figure}
\begin{figure}[h]
\centerline{\includegraphics[width=8cm]{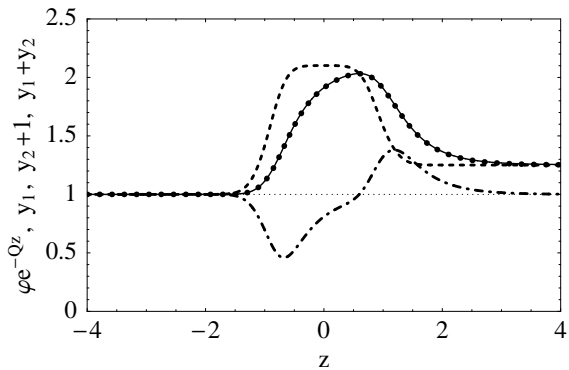}}
\caption{The solution obtained by solving the second order differential equation (full line, equation (\ref{eq:determinantphi}) rescaled as in the figure \ref{fig:seconddiff}) directly, as opposed to the solution of the system of the coupled first order differential equations  in equation (\ref{eq:splitdet}). The dashed line shows the first component of the coupled system, the dashed-dotted line shows the second component plus one, with the circles representing a direct sum of the two solutions. The thin dotted line represents the value of $1$. The two complete solutions are indistinguishable from each other.\label{fig:coupled}}
\end{figure}

In order to demonstrate the concepts outlined in the previous section we introduce a test case of the dielectric profile at a certain imaginary frequency in the integral  (\ref{eq:freeen}), given by
\beq
\left(\ln \epsilon\right)^\prime=A\left[\exp\left(-\frac{(z-a)^2}{2 \sigma^2}\right)-\exp\left(-\frac{(z+a)^2}{2 \sigma^2}\right)\right]
\label{enacba-28}
\eeq
as shown in the figure \ref{fig:diel} with parameters $a=1$, $\sigma=1/4$ and $A=2$, which are also the chosen parameters in all the calculations shown further. 
In the figure \ref{fig:seconddiff} we show the numerical solutions of the differential equations  (\ref{eq:hperp}) and (\ref{eq:determinantphi}) that differ in the sign of the first order derivative term. All calculations are done for $Q=1$. What is shown for each of the differential equations is the ratio of the solution starting with rather arbitrary initial conditions at large negative $z$ (ideally $z\to - \infty$) to the solution of the same differential equation with the same initial condition but for empty space, which itself behaves as $\propto \exp(Qz)$. While the rescaled solutions of the two differential equations can be seen to deviate markedly from one another in the region of the varying dielectric constant as well as in the region of the order of $\propto 1/Q$ beyond it, when $z\to \infty$ they relax to the same value, and for the calculation of the free energy (\ref{eq:freeen}) this value is the only important quantity.
In figure \ref{fig:coupled} the comparison of the direct solution of the differential equation (\ref{eq:determinantphi}) with the solution obtained by splitting it into the system of differential equations (\ref{eq:splitdet}) is shown. Again, the results are divided by the empty space solution $\propto \exp(Qz)$. The sum of the two components of the coupled system of equations is indistinguishable from the direct solution to the original equation. What is apparent is that the solution for the first component of the coupled system of equations varies only in the region of the varying dielectric constant (see the figure \ref{fig:diel}); beyond that region only the second component of the system is still relaxing to $0$ with the first component remaining constant. For the purposes of the calculation of the free energy, the use of a coupled system is therefore beneficial as the calculation can be terminated immediately after entering the homogeneous region, seeing that the first component of the system carries all the information about the required asymptotic behaviour, whereas in the direct solution to the differential equation it is necessary to go deeper into the homogeneous region for the solution to relax.

This shows that using the matrix product formulation based on the field approach has several advantages over the DFT approach in the numerical implementation of the calculation. The main benefit is that the domain of calculation necessary to determine the solution is only that part of the $z$-axis beyond which the dielectric response is constant, whereas in the continuous DFT approach one needs to extend the system by the length $\propto 1/Q$ in order to allow for the relaxation of the solution. Furthermore, the matrix product approach allows for direct treatment of discontinuous dielectric profiles which, while {\sl in principle} accessible {\sl via} the DFT approach, nevertheless require a somewhat special treatment. The advantage of the field approach is particularly relevant for cases where there are several discontinuities or mixed continuous regions interspersed with discontinuities in the dielectric response function that can all be handled quite straightforwardly  in the field approach. Finally, while we showed the equivalence of the two approaches only for the nonretared case of the Van der Waals interactions, valid at small separations between interacting systems, the field approach can be generalized directly and straightforwardly  to calculate the retarded van der Waals - dispersion interactions as well. As far as we know the DFT approach has not yet been generalized into the retarded van der Waals region.

\begin{acknowledgements}
This work has been supported by the European Commission under Contract No. NMP3-CT-2005-013862 (INCEMS).
\end{acknowledgements}

%%\begin{figure}
%%\centerline{\includegraphics[width=13cm]{plot_diag_nodiag.eps}}
%%\caption{The solutions for the first component as obtained by the method of coupled differential
%%equations, but with and without using the diagonal transformation. The purple curve gives the
%% solution with the negative sign and the blue curve is the solution with the positive sign, both without
%% using the diagonal transform. When the diagonal transform is taken into account, solutions for both
%% signs become equal (green, red lines) \label{fig:diagonal}}
%%\end{figure}


\begin{thebibliography}{99}

\bibitem{parsegian} Parsegian VA, {\sl Van der Waals Forces : A Handbook for Biologists, Chemists, Engineers, and Physicists} (Cambridge University Press, 2005).

\bibitem{rudi1} Podgornik R, French RH and Parsegian VA,  {\sl J. Chem. Phys.} {\bf 124} (2006) 044709 .

\bibitem{decher1} Decher G, Eckle M, Schmitt J, Struth B, 
\textsl{Curr Op Colloid Interface Sci}  \textbf{3 } (1998) 32-39. 

\bibitem{sukhishvili} Sukhishvili SA, Granick S, \textsl{Macromol} \textbf{35}  (2002) 301-310. 

\bibitem{roger-1}  French RH,  \textsl{JACS} \textbf{83}  (2000) 2117-2146.   

\bibitem{roger-2}  van Benthem K {\sl et al.}, {\sl Phys. Rev. B} {\bf 74} (2006) 205110.

\bibitem{ninham} Mahanty J and Ninham BW, {\sl Dispersion forces} (Academic
Press, New York) 1976.  Langbein D, {\sl Van der Waals Attraction} (Springer Tracts in Modern Physics, Springer-Verlag, Berlin 1974). Mostepanenko V and   Trunov NN, {\sl The Casimir Effect and Its Applications} (Oxford University Press, USA) 1997.

\bibitem{parr} Parr RG, Yang W, {\sl Density-Functional Theory of Atoms and Molecules} (International Series of Monographs on Chemistry, Oxford University Press, USA; Reprint edition,1994).

\bibitem{rydberg-dion} Rydberg H, Lundqvist BI, Langreth DC and Dion M, {\sl Phys. Rev. B} {\bf 62} (2000) 6997-7006.

\bibitem{langreth} Langreth DC {\sl et al.}, {\sl International Journal of Quantum Chemistry} {\bf 101} (2005) 599-610.

\bibitem{rydberg-dion2} Rydberg H {\sl et al.}, {\sl Phys. Rev. Lett.} {\bf 91} (2003) 126042-1.

\bibitem{dobson} J.F. Dobson and Jun Wang, {\sl Phys. Rev. Letts.} {\bf 82} (1999) 2123-2126.

\bibitem{reform} Podgornik R, Hansen PL and Parsegian VA, {\sl J. Chem. Phys.} {\bf 119} (2003) 1070-1077.


\end{thebibliography}
\end{document}